\documentclass[aps,prl,twocolumn,showpacs,amsmath,amssymb]{revtex4}
\usepackage{epsfig}
\usepackage{bm}

\newlength{\figurewidth}

\newcommand{\be}{\begin{equation}} \newcommand{\ee}{\end{equation}}

\newcommand{\ba}{\begin{eqnarray}}
\newcommand{\ea}{\end{eqnarray}}

\newcommand{\bq}{\begin{equation}}
\newcommand{\eq}{\end{equation}}
\newcommand{\bqa}{\begin{eqnarray}}
\newcommand{\eqa}{\end{eqnarray}}
\newcommand{\ben}{\begin{enumerate}}
\newcommand{\een}{\end{enumerate}}
\newcommand{\bc}{\begin{center}}
\newcommand{\ec}{\end{center}}
\newcommand{\bqb}{\begin{eqnarray*}}
\newcommand{\eqb}{\end{eqnarray*}}

\newcommand{\pap}[4]{#1 {\bf #2}, #3 (#4)}

\begin{document}

\setlength{\figurewidth}{\columnwidth}

\title{\vspace{1cm}
Density Matrix Renormalization Group Study of the \\
Disorder Line in the Quantum ANNNI Model
}

\author{M. Beccaria${}^a$, M. Campostrini${}^b$, A. Feo${}^c$}

\affiliation{
${}^a$ Dipartimento di Fisica and INFN, Universit\`a di
Lecce \\
Via Arnesano, 73100 Lecce, Italy\\
${}^b$ INFN, Sezione di Pisa, and
Dipartimento di Fisica ``Enrico Fermi'' dell'Universit\`a di Pisa,
Largo Bruno Pontecorvo 3, I-56127 Pisa, Italy \\ 
${}^c$ Dipartimento di Fisica, Universit\`a di Parma and INFN Gruppo Collegato di Parma,
Parco Area delle Scienze, 7/A, 43100 Parma, Italy
}

\begin{abstract}
We apply Density Matrix Renormalization Group methods to study the phase diagram of the quantum ANNNI model
in the region of low frustration where the ferromagnetic coupling is larger than the next-nearest-neighbor antiferromagnetic
one. By Finite Size Scaling on lattices with up to 80 sites we locate precisely the transition line from the 
ferromagnetic phase to a paramagnetic phase without spatial modulation. We then measure and analyze the spin-spin
correlation function in order to determine the disorder transition line where a modulation appears.
We give strong numerical support to the conjecture that the Peschel-Emery one-dimensional line actually
coincides with the disorder line. We also show that the critical exponent governing the vanishing of the modulation 
parameter at the disorder transition is $\beta_q = 1/2$. 
\end{abstract}


\maketitle

The ANNNI model is an axial Ising model with competing next-nearest-neighbor antiferromagnetic coupling in one direction.
It is a paradigm for the study of competition between magnetic ordering, frustration and thermal disordering effects.
Its phase diagram displays indeed a rich variety of phases. In the most realistic three-dimensional case, it 
describes several physical systems from  magnetic materials like CeSb to binary alloys or dielectrics
like NaNO${}_2$~\cite{Selke1}.
In the more academic one-dimensional case, it is exactly solvable and several general properties
can be rigorously proved about its phase diagram~\cite{1d}.

The two-dimensional case is nontrivial and not solvable, but its phase structure 
is much simpler than the 3d case. The model is believed to display 5 phases~\cite{5phases}:
Ferromagnetic $\uparrow\uparrow\uparrow\uparrow$, antiphase $\uparrow\uparrow\downarrow\downarrow$, 
paramagnetic with or without modulation and floating phase with algebraically decaying spin correlations.
This picture is supported by various analytical~\cite{analytical}  and numerical~\cite{numerical} 
investigations based on a variety of approximations.
However, lacking an exact solution, the precise location of the 
various transitions is not known beyond approximate treatments. Actually there is no rigorous proof
of the existance of all the above phases. In particular the very existance of the floating phase has been
recently under debate~\cite{Shirahata}.

To further simplify the analysis, the 2d case can be studied in the Hamiltonian limit
which is a one-dimensional quantum spin $S=1/2$ chain with next-nearest-neighbor coupling. 
The chain interacts 
with an external field playing the role of the temperature and triggering phase transitions. 
The Hamiltonian limit, also called TAM model (Transverse ANNNI) is very interesting in 
itself being a simple example where several complicated quantum phase transitions do occur
with drastic changes in the qualitative features of the ground state~\cite{QPT}.

The accurate numerical study of the TAM model is challenging notwithstanding its
relative simplicity. In this Letter we address an open conjecture concerning its
disorder line by employing Density Matrix Renormalization Group methods.

To illustrate the problem, we introdude the TAM Hamiltonian with open boundary conditions
which reads
\be
\label{eq:TAM}
H = -J_1 \sum_{i=1}^{L-1} \sigma_i^z\sigma_{i+1}^z-J_2\sum_{i=1}^{L-2} \sigma_i^z\sigma_{i+2}^z-B \sum_{i=1}^L \sigma_i^x .
\ee
We shall present our results in terms of the adimensional ratios $\kappa = -J_2/J_1$ and $B/J_1$ which are the 
only parameters that describe the properties of the ground state.

The qualitative phase diagram of the TAM model is quite different in the two
regions $\kappa < 1/2$ or $\kappa > 1/2$. For $\kappa < 1/2$, the model is in a ferromagnetic phase
at low magnetic field $B$. At $B_{c, 1}$, a transition in the Ising class makes the ground state 
paramagnetic with exponentially decaying spin-spin correlation functions. Increasing further the external field
we expect a new transition for $B > B_{c, 2} > B_{c, 1}$ where the model is still gapped but with a 
correlation function whose exponential decay has also a spatial modulation. In this phase
the asymptotic correlation function in the bulk is conveniently parametrized for large spin separation  $d$ 
by the functional form
\be
\label{eq:correlation}
C^{zz}(d) = \langle \sigma^z_i\sigma^z_{i+d}\rangle \sim c_0 e^{-d/r} \cos(\pi q d + \varphi),
\ee
with $r$ and $q$ being functions of $B$ and $\kappa$.

The modulation parameter $q(B, \kappa)$ vanishes at $B=B_{c,2}(\kappa)$ 
with a certain exponent $\beta_q$
\be
q(B, \kappa) \sim A(B-B_{c, 2}(\kappa))^{\beta_q},\ \mbox{as}\ B\searrow B_{c, 2}(\kappa).
\ee
The critical line $B = B_{c, 2}(\kappa)$ is known as a disorder line~\cite{1d,DisorderLine} (see also~\cite{Selke2}
for a different definition).

The region $\kappa>1/2$ is much more complicated. At low $B$ the ground state is in a 
so-called antiphase with typical spin configuration $\uparrow\uparrow\downarrow\downarrow\cdots$.
Increasing the magnetic field one expects to observe a first transition to a disordered phase
with algebraically decaying $C^{zz}$ (the floating phase) followed by a final transition to the asymptotic
paramagnetic phase, i.e., the unique high temperature phase in the original 2d statistical model.
The numerical data in this region are controversial and the size  of the floating phase is not clear
being possibly zero~\cite{Shirahata}. 

In this Letter, we shall be concerned with the $\kappa < 1/2$ region only. 
For simplicity, we shall denote this region by LFR (low frustration region).
In the LFR, there is general consensus about the phase diagram, although only at the
qualitative level, i.e., with large variations due to the various approximation employed in 
its calculation.

Remarkably, the TAM model can be solved exactly on a critical line in the LFR
called the Peschel-Emery one-dimensional line (ODL)~\cite{PE}. The spin correlation decays exponentially
on the ODL which is immersed in the paramagnetic phase. Little is known analytically off the ODL
due the the very tricky nature of the solution.
It is still a conjecture that the ODL is indeed the disorder line and that therefore
\be
B_{c, 2}/J_1 = B_c^{PE}/J_1\equiv \kappa-\frac{1}{4\kappa}.
\ee
The conjecture is compatible with the numerical simulations of the TAM model. However, 
the agreement $B_{c, 2} = B_c^{PE}$ is valid at  not more than about 20 \% accuracy along the line.

The aim of this Letter is precisely to give a numerical {\em proof} with good accuracy of 
this conjecture. As a byproduct we also determine the unknown exponent $\beta_q$.

A detailed analysis of the quantum ANNNI model can be found in~\cite{oldANNNI}. The 
accuracy of the results is poor because of the small considered lattices with less than 10 sites.
Another interesting approach is described in~\cite{Rieger} where an effective Hamiltonian is proposed
allowing a considerable reduction of the Hilbert space. Systems up to 32 sites long can be treated, but 
the approximation is valid only near $\kappa = 1/2$.

A more recent numerical analysis of the LFR is~\cite{Colares} where the ferromagnetic-paramagnetic Ising
transition is analyzed by combining Finite Size Scaling with exact diagonalization of short chains with no more than 10 sites. 

Here we present a study of the model with higher accuracy and much  larger lattices by means of the 
Density Matrix Renormalization Group (DMRG) algorithm~\cite{DMRG}.
Nowadays, this method appears to be the natural choice for one-dimensional quantum spin chains.

We have implemented the finite size version of the DMRG algorithm 
computing the two lowest levels $E_{0,1}$ and the energy gap $\Delta=E_1-E_0$.
The algorithm results are very stable when more than 80 states are kept in the block Hamiltonians.
In practice the numerical error on $\Delta$ is at the level of the machine precision.

\begin{figure}[tb]
\vskip 0.7cm
\begin{center}
\leavevmode
\psfig{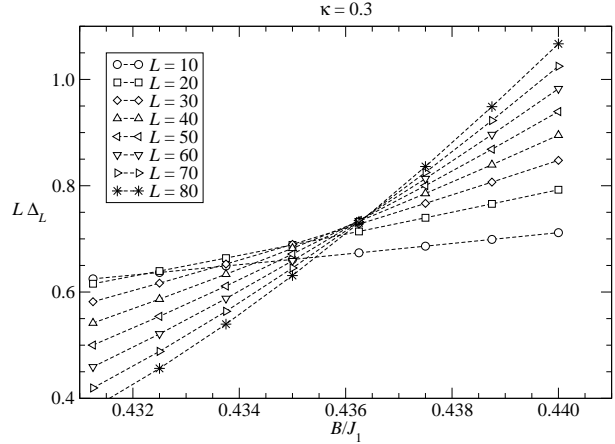}
\vspace{0.1cm}
\caption{Finite Size Scaling analysis of $L\Delta_L(\kappa, B)$ at $\kappa = 0.3$.}
\label{fig:1}
\end{center}
\end{figure}

\begin{figure}[tb]
\vskip 0.7cm
\begin{center}
\leavevmode
\psfig{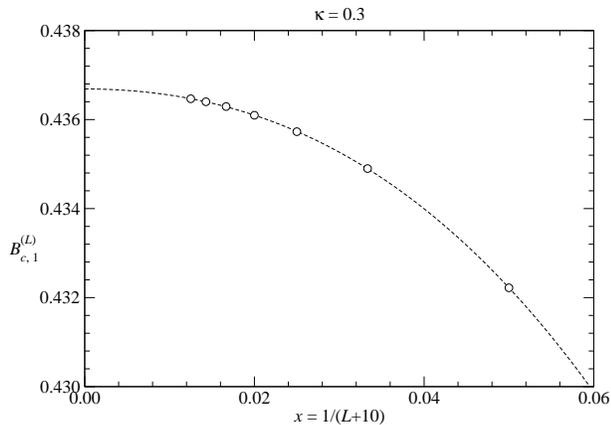}
\vspace{0.1cm}
\caption{Extrapolation of the Finite Size Scaling crossing estimator $B_{c, 1}^{(L)}$. $\kappa = 0.3$ and $L = 10$, $20$, \dots, $80$.}
\label{fig:2}
\end{center}
\end{figure}

\begin{figure}[tb]
\vskip 0.7cm
\begin{center}
\leavevmode
\psfig{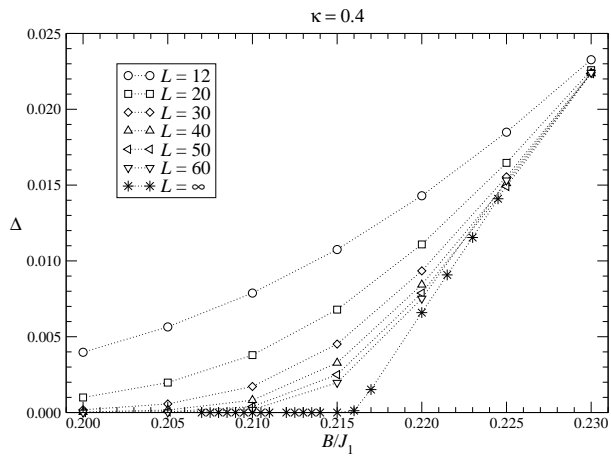}
\vspace{0.1cm}
\caption{Finite Size Scaling analysis at $\kappa = 0.4$. The left plot includes the data obtained with the 
infinite size DMRG algorithm and labeled $L=\infty$.}
\label{fig:2a}
\end{center}
\end{figure}

For several lattice sizes $L$ of order $10^2$ and various frustration ratios $\kappa$, we have computed the scaled energy gap 
$L\Delta_L(\kappa, B)$. The crossing of the associated curves as a function of $B$ at fixed $\kappa$ 
is a finite size estimate of the ferromagnetic critical field $B_{c, 1}^{(L)}(\kappa)$. 
As an example, we show in Fig.~(\ref{fig:1}) our results at $\kappa = 0.3$.
\noindent
We have determined the crossing point between the curves associated to a certain $L$ and $L+10$.
We expect $B_{c, 1}^{(L)}\to B_{c,1}(\kappa)$ as $L\to\infty$ with algebraic corrections in $1/L$~\cite{Hamer}. 
We show in Fig.~(\ref{fig:2}) the finite size crossing field plotted as a function of $x=1/(L+10)$
which is the most convenient variable to extrapolate our $(L, L+10)$ crossings. Indeed, the 
fitting function $a+b x^2 + c x^3$ gives a very good $\chi^2$ of about $10^{-11}$.
\noindent
The results for all the  considered $\kappa$  are collected  in Tab.~(\ref{tab:1})
where we also show (when available) the analogous results from~\cite{Colares}. These are obtained
on small lattices crossing $L$ with a fixed $L=4$. This is at most an estimate of $B_{c, 1}$.
Tab.~(\ref{tab:1}) reports also $\widetilde{B}_{c, 1}$ which is obtained from the 
vanishing of the gap at second order in $B$~\cite{PE}, and is defined by 
\be
1+2\kappa = \frac{\widetilde{B}_{c, 1}}{J_1} + \frac{\kappa}{2(1+\kappa)}\left(\frac{\widetilde{B}_{c, 1}}{J_1}\right)^{\!\! 2}.
\ee
In principle, it is possible to determine $B_{c, 1}$ directly in the infinite size limit by using the infinite
lattice version of the DMRG algorithm. We show in Fig.~(\ref{fig:2a}) the result of such a procedure at $\kappa = 0.4$.
The result for $B_{c, 1}$ is fully consistent with the FSS analysis. Also, the exponent $\nu=1$ which 
governs the vanishing of the mass gap is very clear at $L=\infty$. For the other values of $\kappa$ we have preferred to avoid the 
infinite size algorithm since it is known that it can fail when the phase structure is complicated~\cite{DMRGfail}.
\noindent
From Tab.~(\ref{tab:1}) we see that the DMRG estimate $B_{c, 1}^{\rm DMRG}$ and the 
results of~\cite{Colares} are globally similar and slightly below the approximation $\widetilde{B}_{c, 1}$,
especially at large $|\kappa-1/2|$. The value from~\cite{Colares} at $\kappa = 0.4$ is somewhat away from the 
common values of $B_{c, 1}^{\rm DMRG}$ and $\widetilde{B}_{c, 1}$.
\begin{table}
\begin{tabular}{l|lll|ll}
$\kappa$ & $B_{c, 1}^{\rm DMRG}/J_1$ & $B_{c, 1}/J_1$ \cite{Colares} & $\widetilde{B}_{c, 1}/J_1$ & $B_{c, 2}^{\rm DMRG}/J_1$ & $B_{c, 2}/J_1$ \cite{PE} \\
\hline
0.15 & 0.73405(4)  & 0.7327(2)  & 0.74956 & 1.5168(2)  & 1.51667 \\ 
0.20 & 0.6393(1)   & 0.6407(4)  & 0.65336 & 1.0500(1)  & 1.05    \\
0.25 & 0.5403(3)   & 0.5388(4)  & 0.55051 & 0.75001(2) & 0.75    \\
0.30 & 0.43669(4)  & 0.4368(2)  & 0.44183 & 0.53337(5) & 0.53333 \\
0.35 & 0.32821(2)  & 0.3298(3)  & 0.32917 & 0.36428(5) & 0.36429 \\
0.40 & 0.216090(3) & 0.2068(3)  & 0.21548 & 0.22498(2) & 0.225 
\end{tabular}
\caption{Comparison of the ferromagnetic and disorder critical fields $B_{c, 1-2}/J_1$.
The DMRG columns are our data. $\widetilde{B}_{c, 1}$ is obtained from the vanishing of the gap at
second order in the magnetic field. The data from~\cite{Colares} have been interpolated at $\kappa = 0.15$, 0.35.}
\label{tab:1}
\end{table}

\begin{figure}[tb]
\vskip 0.7cm
\begin{center}
\leavevmode
\psfig{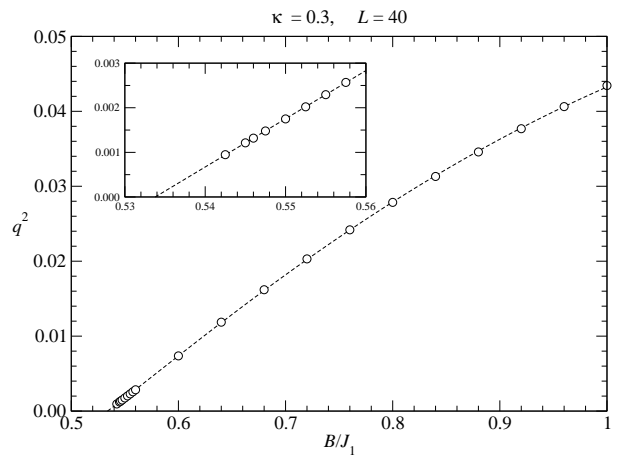}
\vspace{0.1cm}
\caption{$B$ dependence of the squared modulation parameter $q^2$ at $\kappa = 0.3$. The fit is performed on the 
leftmost points near the critical point.}
\label{fig:3}
\end{center}
\end{figure}

After the determination of the ferromagnetic-paramagnetic Ising transition, 
we studied $B_{c, 2}(\kappa)$ and the critical behavior of the modulation $q$.
We have computed by the DMRG algorithm the spin correlation $C^{zz}$ on lattices large compared to the 
correlation length $r$ appearing in Eq.~(\ref{eq:correlation}). In practice, $L=40$ is enough in all the considered cases.
The critical behaviour of $q(B)$ is shown in Fig.~(\ref{fig:3}) for $\kappa = 0.3$. The vanishing of $q^2$
is linear in $B-B_{c, 2}$. 
\noindent
The modulation parameter vanishes with exponent $\beta_q = 1/2$. The critical field $B_{c, 2}$
coincides with the Peschel-Emery value~\cite{PE} with high accuracy.

We have repeated the analysis for $\kappa = 0.15$, 0.2, 0.25, 0.35, 0.4, finding always 
a very good agreement. The agreement at small frustration is remarkable. 
From the point of view of the disorder line the next-nearest-neighbor coupling 
$J_2$ is a singular perturbation with $B_{c, 2}\to\infty$ in the isotropic Ising limit
$\kappa\to 0$. We remark that it is nontrivial to 
extend the calculation of~\cite{PE} off the one-dimensional line to proof rigorously that the 
one-dimensional line is the disorder line. Indeed, the only analytic insight in this direction 
is the analysis in~\cite{SenLargeS} where the ODL is proved to
be the disorder line, but only mapping the initial $S=1/2$ spin chain into a dual 
spin $T=1/2$ chain and taking the $T\to \infty$ limit.

\begin{figure}[tb]
\vskip 0.7cm
\begin{center}
\leavevmode
\psfig{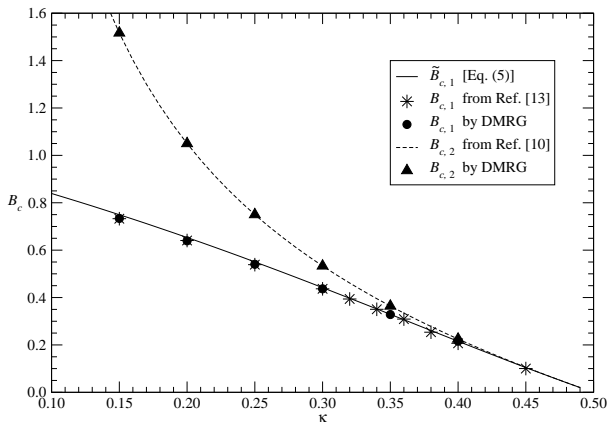}
\vspace{0.1cm}
\caption{Phase diagram in the LFR showing the agreement between the disorder line computed with DMRG
and the exact one-dimensional line by Peschel-Emery. See Tab.~(\ref{tab:1}) for numerical values.}
\label{fig:4}
\end{center}
\end{figure}

Our results for $B_{c, 2}$ are also summarized in Tab.~(\ref{tab:1}) together with the Peschel-Emery
value. In Fig.~(\ref{fig:4}) we plot the final phase diagram as determined by our DMRG simulations.

In conclusion, we have shown that a DMRG analysis of the quantum ANNNI model 
provides strong numerical support to the conjecture that the Peschel-Emery ODL is 
actually the disorder line. Also, the critical exponent
governing the vanishing of the modulation at the disorder transition
is $\beta_q = 1/2$.

A natural extension of this work concerns the DMRG study of the region $\kappa > 1/2$,
which requires much larger lattices to analyze the slow algebraic decay of the spin correlation 
functions in the would-be floating phase. 

We acknowledge conversations with G.~F.~De Angelis, W.~Selke,
P.~Fendley, and V.~Rittenberg.

\begin{thebibliography}{99}

\bibitem{Selke1}
An old review is: W. Selke,
\pap{Phys. Rep.}{170}{213}{1988}. More recent experimental results concerning 
incommensurate stripe structures in various materials can be found for
instance in 
S. Kawano and N. Achiwa, \pap{J. Magn. Magn. Mater.}{ 52}{ 464}{1985};
X. Dai, Z. Xu, and D. Viehland, \pap{J. Am. Ceram. Soc. }{78}{2815}{1995};
J. Ricote et al., \pap{J. Phys.: Condens. Matter}{10}{ 1767}{1998};
D. Viehland, \pap{Phys. Rev. }{B 52}{ 778}{1995};
V. Massidda and C.R. Mirasso, \pap{Phys. Rev. }{B 40}{9327}{1989};
D. Bitko, T.F. Rosenbaum, and G. Aeppli, \pap{Phys. Rev. Lett. }{77}{ 940}{1996}.

\bibitem{1d}
R. Liebmann, Lecture Notes in Physics 251 (Springer Berlin 1986);
J. Stephenson, \pap{Can. J. Phys.}{48}{1724}{1970};
T. Oguchi, \pap{J. Phys. Soc. Jpn.}{20}{2236}{1965};
R. M. Hornreich, R. Liebmann, H. G. Schuster and W. Selke, \pap{Z. Phys. }{B35}{91}{1979}.

\bibitem{5phases}
See for instance 
D. Allen et al.,
\pap{J. Phys. A: Math. Gen.}{34}{L305-L310}{2001}.

\bibitem{analytical}
E. M\"uller-Hartmann and J. Zittartz, \pap{Z. Phys. }{B27}{261}{1977};
J. Villain and P. Bak, \pap{J. Phys. (Paris)}{ 42}{657}{1981};
J. Kroemer and W. Pesch, \pap{J. Phys. }{A 15}{L25}{1982};
M. N. Barber, P.M. Duxbury, \pap{J. Phys.}{A14}{L251}{1981}, \pap{J. Phys.}{A15}{3219}{1982};
M. D. Grynberg and H. Ceva, \pap{Phys. Rev. }{B 36}{ 7091}{1987};
M.A.S. Saqi and D.S. McKenzie, \pap{J. Phys. }{A 20}{ 471}{1987};
Y. Murai, K. Tanaka, and T. Morita, \pap{Physica}{A 217}{214}{1995}.

\bibitem{numerical}
W. Selke and M. E. Fisher, \pap{Z. Phys. }{B 40}{ 71}{1980};
W. Selke, \pap{Z. Phys.}{B 43}{ 335}{1981};
A. Sato and F. Matsubara, \pap{Phys. Rev.}{B 60}{ 10 316}{1999}.

\bibitem{Shirahata}
T. Shirahata and T. Nakamura,
\pap{Phys. Rev.}{ B 65}{ 024402}{2002}.

\bibitem{QPT}
See for instance D. V. Shopova, D. Uzunov,
\pap{Phys. Rep.}{379}{1}{2003} or the recent book 
S. Sachdev, {\em Quantum Phase Transitions},
ISBN: 0521004543, Cambridge University Press, (2001).

\bibitem{DisorderLine}
J. Stephenson, \pap{Phys. Rev.}{B1}{4405}{1970}.

\bibitem{Selke2}
K. Binder, W. Kinzel, W.  Selke, \pap{Surf. Science}{125}{74}{1983}.

\bibitem{PE}
I. Peschel, V. J. Emery, \pap{Z. Phys.}{B43}{241}{1981}.

\bibitem{oldANNNI}
C. M. Arizmendi, A. H. Rizzo, L. N. Epele and C. A. Garcia Canal,
\pap{Eur. Phys. J.}{ B 83}{273}{1991};
P. Sen, S. Chakraborty, S. Dasgupta, B. K. Chakrabarti,
\pap{Z. Phys. }{B 88}{333}{1992}.

\bibitem{Rieger}
H. Rieger, G. Uimin,
\pap{Z. Phys.}{ B 101}{597}{1996}.

\bibitem{Colares}
P. R. Colares Guimares, J. A. Plascak, F. C. S. Barreto, J. Florencio,
\pap{Phys. Rev.}{B 66}{064413}{2002}.

\bibitem{DMRG}
S. R. White, \pap{Phys. Rev. Lett.}{69}{2863}{1992};
\pap{Phys. Rev.}{ B 48}{10345}{1993}.
A complete updated bibliography of applications of the DMRG can be found in the 
recent review 
U. Schollwoeck, \pap{Rev. Mod. Phys.}{77}{259}{2005},
and at the web site
http://quattro.phys.sci.kobe-u.ac.jp/dmrg.html.

\bibitem{Hamer} 
C J Hamer et al.,
\pap{ J. Phys. A: Math. Gen.}{14}{}{ 1981}.

\bibitem{DMRGfail}
U. Schollw\"ock, S. Chakravarty, J. O. Fj\"orestad, J. B. Marston, and M. Troyer,
\pap{Phys. Rev. Lett.}{90}{186401}{2003}.

\bibitem{SenLargeS}
D. Sen,
\pap{Phys. Rev.}{ B 43}{ 5939}{1991}.


\end{thebibliography}
\end{document}